\documentclass[aps,prb, reprint, amsmath, amssymb, groupedaddress, superscriptaddress, showkeys,floatfix]{revtex4-1}

\usepackage{hyperref}
\usepackage{graphicx}
\usepackage{dcolumn}
\usepackage{bm}

\usepackage{amsmath}
\usepackage{amssymb}
\usepackage[usenames]{xcolor}
\usepackage[normalem]{ulem}
\usepackage{soul}
\setulcolor{red}
\usepackage[english]{babel}

\def\Gammapoint{${\bar\Gamma}$}
\def\GammaK{${\bar\Gamma}-{\bar{\rm K}}$}
\def\GammaM{${\bar\Gamma}-{\bar{\rm M}}$}

\def\KGammaM{$\bar{\mathrm K}-\bar\Gamma-\bar{\mathrm M}$}

\begin{document}

\title{Interplay between surface Dirac and Rashba states specific for topologically nontrivial van der Waals superlattices}

\author{I. A. Shvets}
 \email{ shvets\_ia@mail.ru}
 \affiliation{Tomsk State University, 634050 Tomsk, Russia}

\author{E. V. Chulkov}
\affiliation{Departamento de Pol\'imeros y Materiales Avanzados: F\'isica, Qu\'imica y Tecnolog\'ia, Facultad de Ciencias Qu\'imicas, Universidad del Pa\'is Vasco UPV/EHU, 20080 San Sebasti\'an/Donostia, Spain}
\affiliation{Donostia International Physics Center (DIPC), 20018 Donostia-San Sebasti\'{a}n, Basque Country, Spain}
\affiliation{Centro de F\'{i}sica de Materiales (CFM-MPC), Centro Mixto CSIC-UPV/EHU,  20018 Donostia-San Sebasti\'{a}n, Basque Country, Spain}
\affiliation{Saint Petersburg State University, 198504 Saint Petersburg, Russia}

\author{S. V. Eremeev}
\affiliation{Institute of Strength Physics and Materials Science, Russian Academy of Sciences, 634055 Tomsk, Russia}

\date{\today}

\begin{abstract}
Here we show that, in contrast to the observed surface states in well studied pnictogen chalcogenide van der Waals (vdW) topological insulators (TIs) with quintuple layer (QL) or septuple layer~(SL) structure, in superlattices, comprising the alternating QL and SL vdW blocks, the Dirac state becomes accompanied by emergent spin-polarized states of the Rashba type. This specific feature is caused by an inequivalence of the surface and subsurface structural blocks and an electrostatic potential bending near the surface. Within density functional theory~(DFT) and $\emph{ab-initio}$ tight-binding~(TB) calculations we analyze peculiarities of these states depending on the surface termination, structural parameters and chemical composition. It is found that their possible hybridization with the Dirac state significantly affects its dispersion and spatial localization. We analyze the influence of intrinsic magnetism on behavior of the termination-dependent surface states for magnetic QL/SL superlattices. These findings provide a better understanding of the existing experimental observations of such QL/SL alternating superlattices.
\end{abstract}

\maketitle

\section{Introduction}\label{section_1}

A three-dimensional topological insulator (3D TI) is characterized by surface states arising due to a strong spin-orbit coupling and being protected by a time-reversal symmetry~\cite{qi2008topological}. Bismuth chalcogenides, Bi$_2$Te$_3$ and Bi$_2$Se$_3$, are among the first 3D TIs composed of QL vdW blocks~\cite{zhang2009topological} and are still attracting the utmost interest. The practical requirements to 3D TIs motivated a search and synthesis of new materials with more suitable electronic structure parameters and the largest contribution of the surface conductivity realized by topological surface states (TSS) to the total conductivity~\cite{Ando_review}. In this regard, the electronic structure of the vdW materials can be widely manipulated due to structural and compositional variations \cite{eremeev2012effect,forster2015abinitio}. The following search strategies for novel non-magnetic vdW TIs have early been suggested: a complete or partial replacement of certain layers of atoms by isovalent ones -- Bi$\rightarrow$Sb, Te$\rightarrow$Se$\rightarrow$S (Bi$_2$Te$_2$Se~\cite{Ando_review}, Bi$_2$Te$_2$S~\cite{Annese_PRB2018}, Bi$_{2-x}$Sb$_x$Te$_{3-y}$Se$_y$~\cite{filianina2018spin}); an increase in the number of atoms in a unit cell and incorporation of a new sort of atoms -- Pb and Ge, leading to a synthesis of three/four component compounds (GeBi$_2$Te$_4$~\cite{Okamoto2012}, PbBi$_{2}$Te$_4$~\cite{PbBi2Te4_Kuroda}, Sn(Pb)Sb$_2$Te$_4$ \cite{Menshchikova2013}, PbBi$_{2}$Te$_2$Se$_2$~\cite{shvets2017impact} composed of SL vdW blocks), including solid solutions (Pb(Bi$_{1-x}$Sb$_x$)$_2$Te$_4$~\cite{Souma2012}). In view of the possible practical use of heterostructures consisting of trivial and topological materials~\cite{zhang2010crossover, sakamoto2010spectroscopic, taskin2012manifestation, wu2013sudden, kim2013coherent, neupane2014observation, landolt2014spin, jiang2012landau}, superlattices composed of structural blocks of different composition are of a great interest.

A number of structures (PbBi$_{4}$Te$_{7}$, PbBi$_{6}$Te$_{10}$, Sn(Bi,Sb)$_{4}$Te$_{7}$, etc.), formed by alternating QL [$X_{2}$$Y_{3}$] and SL [$ZX_{2}$$Y_{4}$] ($X$~=~Bi, Sb; $Y$~=~Te, Se, S; $Z$~=~Pb, Sn, Ge) blocks, each of which is a building block of the corresponding 3D TI, have been theoretically predicted \cite{eremeev2010possible,eremeev2012atom,Neupane_PRB2012,vergniory2013bulk,vergniory2015electronic} and experimentally studied \cite{papagno2016multiple,okuda2013experimental,eremeev2012atom,pacile2018deep,sumida2018enhanced,grimaldi2020electronic,shvets2019surface}. In such vdW superlattices, when the crystal is cleaved along the vdW plane, terraces with non-equivalent surface terminations may appear. As a result, upon experimental measurements of the specimen surface, several states with linear dispersion resembling the Dirac cone are simultaneously observed in the electronic energy spectra. Apart from the relative shift of the Dirac points in energy, the observed TSSs have other peculiarities distinguishing them from the TSS on the surface of 3D TIs solely composed of QL or SL structural blocks. Namely, within the DFT calculations depending on the surface termination they exhibit a complex dependence of the spatial distribution on the momentum with a large localization length and more intricate dispersion~\cite{eremeev2010possible,eremeev2012atom,vergniory2013bulk,vergniory2015electronic,papagno2016multiple}. In turn, localization of the Dirac state in subsurface blocks allows to argue in favour of a possibility for protection of the "hidden" TSS from {environment effect. On the ARPES spectra of the first grown and experimentally studied QL/SL 3D TI PbBi$_{4}$Te$_{7}$~\cite{eremeev2012atom,okuda2013experimental}, the spectral branches of the upper part of the Dirac cone originated from the QL terraces were characterized by an unusually weak intensity \cite{okuda2013experimental} and a dispersion with a kink along the {\GammaM} direction, which was also confirmed by the latest measurements~\cite{grimaldi2020electronic}. In Ref.~\cite{grimaldi2020electronic}, in addition to PbBi$_{4}$Te$_{7}$, for the isomorphic PbBi$_{4}$Te$_{6}$Se compound with a~central selenium layer in a QL block the similar features of the surface spectrum were demonstrated. However, to date the physical mechanisms responsible for the formation of such intricate surface electronic structure remain unclear. This question becomes especially relevant in the context of magnetic TI superlattices where the electronic structure is more complicated because of the intrinsic magnetism. In particular, in magnetic MnBi$_4$Te$_7$ it was assumed that a certain trivial surface state in the valence band interacts with the TSS near the \Gammapoint-point, leading to the appearance of the "avoided crossing'' gap \cite{klimovskikh2020tunable}.

The extensive studies on MnBi$_2$Te$_4$ family magnetic TIs are mainly
driven by two motivations: realizing a high-temperature quantum anomalous Hall effect (QAHE) and exploring the exotic phases
arising from its topological and magnetic orders, such as the axion insulator phase \cite{Otrokov_2019_Nature,He2020,Zhao2021}. The magnetic superlattices MnBi(Sb)$_4$Te$_7$ composed of alternating MnBi(Sb)$_2$Te$_4$ SLs and Bi(Sb)$_2$Te$_3$ QLs are close structural relatives of non-magnetic QL/SL superlattices. They are the first ($n=1$) members in series (MnBi(Sb)$_2$Te$_4$)$\cdot$(Bi(Sb)$_2$Te$_3$)$_n$ vdW compounds that harbor multiple topologically nontrivial magnetic phases \cite{klimovskikh2020tunable,eremeev2021topological}. Like parent MnBi$_2$Te$_4$ compound \cite{Eremeev.jac2017,Otrokov_2019_Nature} the MnBi$_4$Te$_7$ belongs to intrinsic antiferromagnetic topological insulator (AFMTI) phase \cite{klimovskikh2020tunable}. Despite earlier studies at normal conditions \cite{Zhang.prl2019, Chen.ncomms2019, Zhou.prb2020, Lei.pnas2020} have shown that MnSb$_2$Te$_4$ is a topologically trivial magnetic insulator, later works \cite{Wimmer2021,Huan_PRL2021}  showed that the topological phase can be achieved in both parent MnSb$_2$Te$_4$ and its  QL/SL superlattice MnSb$_4$Te$_7$ in agreement with DFT calculations \cite{eremeev2021topological}. Since Bi(Sb)-based vdW compounds are intrinsically $n$($p$)-doped crystals, the Bi$_{1-x}$Sb$_x$-based magnetic TIs, possessing a charge neutrality point at $x\approx 0.3$, are commonly grown. In general, in Mn(Bi$_{1-x}$Sb$_x$)$_2$Te$_4$-based structures
 the magnetic state can depend on growth conditions which control concentration and distribution of anti-site defects \cite{Huan_PRL2021,Lin2023,Liu_PRX2021,Chen_PRB2021,Xu_NatComm2022,Qian_NanoLett2022}. In particular case of  Mn(Bi$_{1-x}$Sb$_x$)$_4$Te$_7$ superlattice the AFM ground state evolves to a FM state at the same $x = 0.3$ and then returns to an AFM state at $x = 0.36$ \cite{Chen_PRB2021}.

In this paper, within DFT and TB calculations we analyze peculiarities of the surface states in non-magnetic and magnetic QL/SL superlattices depending on the surface termination, structural parameters and chemical composition. 
Though most of the considered non-magnetic materials are hypothetical, they are composed of well-known structural blocks. Such compounds are designed to trace the evolution of the electronic structure in known QL/SL superlattices with changes in atomic composition. We give an explanation of the complex surface electronic structure near the Fermi level featuring a possible hybridization of the Dirac state with a specific Rashba surface state, emerged owing to the Dirac state induced electrostatic potential in the subsurface layer blocks. Obtained results could give a better understanding of the existing experimental electronic structure observations in such QL/SL alternating superlattices. More specifically, the explanation of this phenomenon can be helpful in separating the different effects that take place in the recently discovered magnetic topologically nontrivial vdW materials. As a magnetic QL/SL superlattice we considered Mn(Bi$_{0.7}$Sb$_{0.3}$)$_4$Te$_7$ system in both AFM and FM magnetic states. We also did calculations for the superlattice in AFM state with $x=0.5$ for comparison.

The rest of the paper is organized as follows. In Section 2, the computational details for the \emph{ab initio} calculations of the atomic structure and the electronic band structure are described. In Section 3, we present the electronic and spin structures of the surfaces of non-magnetic vdW superlattices with periodic alternation of QL and SL blocks, where for descriptive purposes we choose QL = Bi$_2$Te$_3$, Bi$_2$Te$_2$Se, Sb$_2$Se$_3$, Sb$_2$S$_3$ and SL$~$=$~$PbBi$_2$Te$_2$Se$_2$, GeSb$_2$S$_4$. We specify the mechanisms of formation of electronic states near the Fermi level and identify their features depending on the surface termination and QL(SL) block composition. Section 4 is devoted to the Mn(Bi$_{0.7}$Sb$_{0.3}$)$_4$Te$_7$ superlattice for which we analyze the influence of magnetism on behavior of the termination-dependent surface states. The main results are summarized in Section 5 along with the concluding remarks.

\section{Methods and computational details}\label{section_2}

Currently synthesized and still hypothetical structures of the QL/SL type considered in this work have a trigonal crystal structure (group P$\overline{3}$m1) formed by alternating QL and SL blocks (Fig.~\ref{fig1}(a)). Neighboring blocks are weakly coupled by vdW forces. For this reason, the crystal cleavage along the (0001) plane occurs along vdW gaps, and QL or SL terraces can appear on the surface.

The electronic structure calculations were carried out within the framework of DFT using the VASP code \cite{kresse1993ab,kresse1996efficiency}. The projected augmented wave (PAW) method was used in the calculation scheme \cite{blochl1994projector,kresse1999ultrasoft}
with generalized gradient approximation (GGA) \cite{perdew1996generalized} for the exchange-correlation potential. Relativistic effects were taken into account, including the spin-orbit interaction. In order to describe the vdW interactions we made use of the DFT-D3 functional with Becke-Johnson dumping scheme~\cite{Grimme2011}. The geometry optimization was performed until the residual forces on atoms became smaller than 1 meV/\AA.
Dealing with Mn(Bi$_{1-x}$Sb$_{x}$)$_4$Te$_7$ the Mn $3d$-states were treated employing the GGA$+U$ approach \cite{Anisimov1991} within the Dudarev scheme \cite{Dudarev.prb1998}. The $U_\text{eff}=U-J$ value for the Mn 3$d$-states was chosen to be equal to 5.34~eV. The Bi-Sb site intermix was simulated by means of the virtual crystal approximation (VCA) \cite{VCA}. Lattice constants and interlayer distances in this case were calculated from equilibrium structures of MnBi$_4$Te$_7$ and MnSb$_4$Te$_7$ in accordance with the Vegard's law.

The following \emph{k}-point grids were used for self-consistent calculations: $7\times7\times7$ for bulk and $10\times10\times1$ for slabs. The surfaces of non-magnetic superlattices were modeled using symmetrical slabs separated by the vacuum gap in the repeated-slab approach with a thickness of 41 and 43 atomic layers for QL and SL terminations, respectively. For magnetic superlattice in AFM phase the thickness of the SL-terminated slab was kept equal to 43 layers while for QL-terminated slab it was enlarged to 53 atomic layers to ensure an even number (four) magnetic SL blocks, and hence a zero total magnetic moment in the AFM phase. For ferromagnetic slab with QL termination the same 53-layer slab was used while for the SL-terminated case the thickness was enlarged to 55 layers. The vacuum gap was chosen to be 15 {\AA} in all considered models.

\emph{Ab-initio} TB model calculation included a wannierization of the DFT-derived bulk wave functions and construction of respective Hamiltonians in the maximally localized Wannier function basis using WANNIER90 code~\cite{pizzi2020wannier90}. Within this approach the surface electronic structures were obtained using the iterative Green's function method for the semi-infinite crystal geometry utilizing WANNIERTOOLS package~\cite{WU2017}.

\section{Electronic states on the surface of the non-magnetic QL/SL superlattices}
 \label{section_3}

We start our analysis with the Bi$_2$Te$_2$Se/PbBi$_2$Te$_2$Se$_2$ 3D superlattice (Fig.~\ref{fig1}(a)), which has already been studied earlier by DFT calculations and ARPES measurements~\cite{shvets2019surface}. The strong spin-orbit coupling (SOC) in this system leads to inversion of the bulk bands near the Fermi level in the vicinity of the A point~(Figs.~\ref{fig1}(b, c)).  Similar to other known QL/SL structures~\cite{eremeev2010possible, eremeev2012atom, vergniory2015electronic, grimaldi2020electronic} the orbitals of atoms belonging to different blocks make a predominant contribution to distinct bands both in the spectrum without and with SOC turned on (see pink and red circles corresponding to the weights of the Bi orbitals originated from the QL and SL blocks).

\begin{figure*}[t]
\begin{center}
\includegraphics[width=0.8\textwidth]{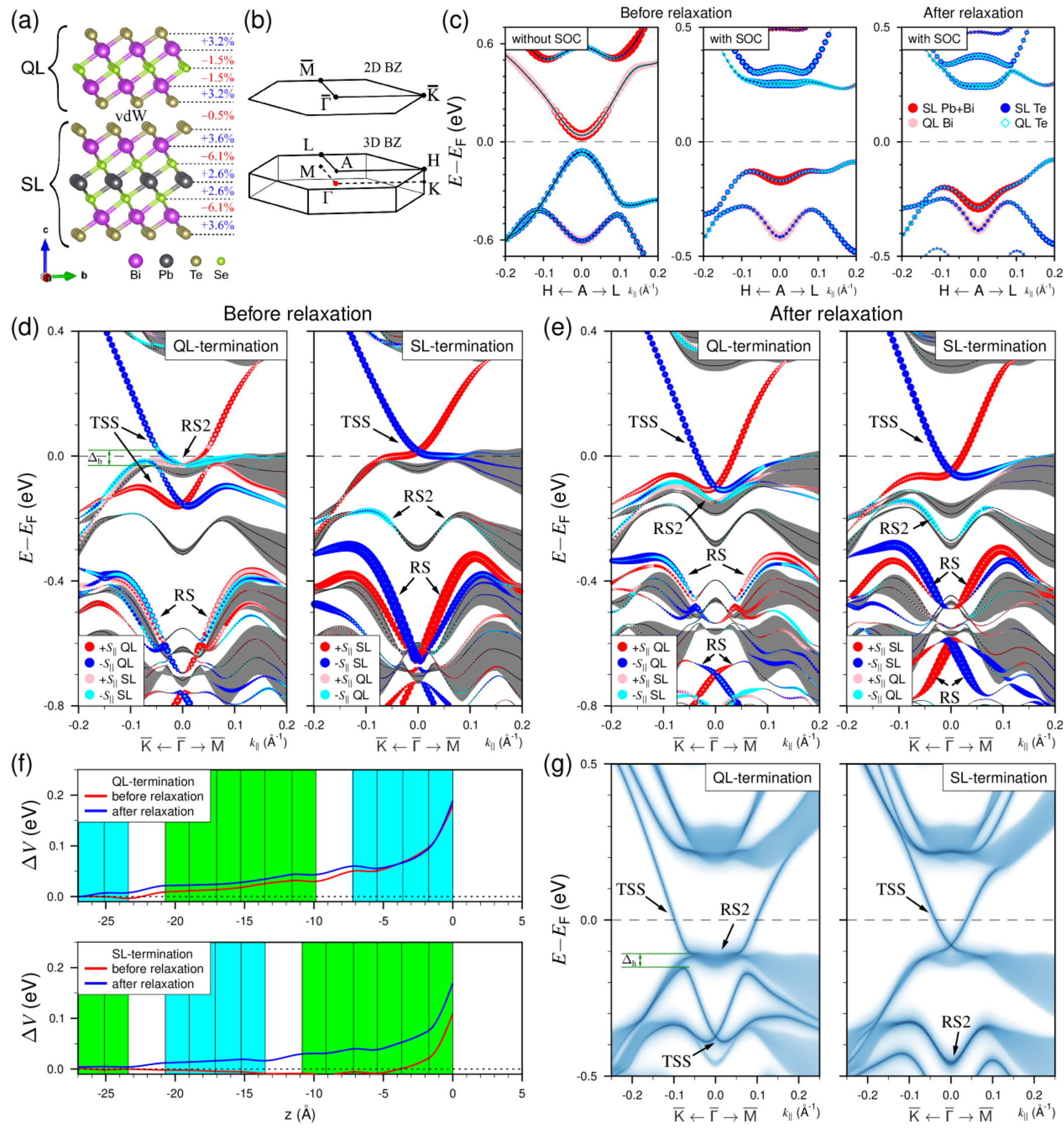}
\caption{(a) Crystal structure of the topologically nontrivial Bi$_2$Te$_2$Se/PbBi$_2$Te$_2$Se$_2$ superlattice, the relative difference between experimental and optimized interlayer distances are shown in percentages. (b) 3D and 2D Brillouin zone schematics. (c) Orbital-resolved bulk band structure, the orbital composition is described by colors shown in the key. (d-e) In-plane spin-resolved band structure of the (0001) slabs along the {\KGammaM} direction with QL and SL block as a surface termination before and after relaxation of the bulk unit cell. Spin structure of the surface states is coded by red (pink) and blue (cyan) for positive and negative value of spin projection of the atom orbitals originated from the first (second) block to the surface. The "TSS", "RS", "RS2" abbreviations accordingly stand for the topological Dirac state, the Rashba states common for 3D TI and the Rashba states intrinsic to the considered TI superlattices, see text. (f) Potential bending $\Delta V$ within the Bi$_2$Te$_2$Se/PbBi$_2$Te$_2$Se$_2$ slab with QL and SL termination before and after the structural relaxation; $z=0$ position corresponds to the surface atomic layer. (g) Tight-binding calculation of surface spectra corresponding to the "before relaxation" (d) case.
}\label{fig1}
\end{center}
\end{figure*}

On the QL surface termination (Fig.~\ref{fig1}(d)) resulted by truncation of the bulk crystal with experimental structural parameters \cite{shvets2019surface} (labeled by "before relaxation"), there can be noticed a spectrum feature near the first bulk valence band. It manifests in the Dirac state ("TSS") piercing through the bulk band and its varying spatial localization preserving the specific for 3D TI spin texture. On the SL termination, however, the spectrum looks like for prototypical 3D TIs -- a single Dirac cone localized primarily in the surface SL block dispersing within the principal band gap.
Another typical surface state derived in theory and observed in experiment \cite{Pauly2012,SEIBEL2015110} is the conventional Rashba-type state at higher binding energy (see "RS" on Figs.~\ref{fig1}(d,e) at $E\sim [-0.7, -0.3]$~eV), which is common for the compounds with strong SOC and 3D TI in particular. In addition, it is worth to mention, that the potential bending on the surface can lead to appearance of a parabolic band in the energy gap just below the conduction band and M-shaped Rashba states in the local gap of the bulk-projected valence band. The emergence of these states is commonly associated with the metallic atoms deposition, effect of residual gases or other inevitable contamination~\cite{Valla2012,Scholz2012,King2011,bahramy2012emergent,eremeev2012effect,Menshchikova2011}, therefore, can be related to external nature. Moreover, the surface steps of TIs can create robust Rashba edge states which interact with the TSSs~\cite{ko2024interplay}. In contrast, as will be shown below, there is another, intrinsic, mechanism for appearing of the Rashba spin-split state manifesting in a different way.

The structural relaxation, performed with vdW force corrections included, leads only to a slight change in the lattice parameters: $a$ by $\sim -0.01$ \AA~($-0.2\%$) and $c$ by $\sim -0.2$~\AA~($-0.8\%$). The relaxation also results in overall small alterations in the interlayer distances (see Fig.~\ref{fig1}(a)), the largest of which is 6.1\% decrease in interlayer distance between Bi and Se layers and 3.6\% expansion between Te and Bi layers in the SL vdW block. Discrepancies of such value between the experimental positions of the layers and those calculated for an ideal equilibrium structure can be observed  in the SL-structured compounds due to the inevitable presence of antisite defects in the samples and slight deviation from stoichiometry \cite{shvets2017impact}.

Relatively small changes in the crystal structure caused by the relaxation lead to a modification of the bulk spectrum (Fig.~\ref{fig1}(c)). Specifically, near the A point the first bulk valence band ($E\sim [- 0.2, -0.1]$ eV), mostly formed by the orbitals of the atoms of the SL blocks, shifts ($E\sim [- 0.3, -0.2]$ eV) towards the second one, which in turn is composed of the orbitals of the QL block atoms and remains not affected by structural changes. This does not cause qualitative variation in the spectrum of the SL-terminated surface (Fig.~\ref{fig1}(e)) as compared with that "before relaxation" case (Fig.~\ref{fig1}(d)). However, there are noticeable changes in the spectrum of the QL-terminated surface, where the most striking difference is that after relaxation the Dirac point ends up in the principal gap. For both terminations the states, predominantly formed by orbitals originated from QL blocks, remain almost unchanged. This can be associated with more pronounced structural relaxation occurring in the SL block. At the same time, it is clearly seen that a spin-split pair of surface states (which negative/positive sign of spin polarization is marked by cyan/pink colors), formed by the orbitals of the subsurface SL block, split off from the top of the bulk band at $E\sim [- 0.2, -0.1]$~eV ("RS2", which hereafter denotes the second vdW block Rashba state). On the energy scale these states are located just below the topological state, localized within the surface QL block, and do not hybridize with it unlike in the "before relaxation" spectrum, where the hybridization gap ($\Delta_{\rm h}$) in TSS is about 50 meV. The dispersion behaviour and the spin structure of these split-off states indicate that they are branches of the Rashba-type state. On the SL termination a pair of similar states can also be seen in the energy range $E\sim [-0.3, -0.2]$ eV,
slightly
split off from the second bulk band formed by the orbitals of the QL blocks.

Thus, in the case of "before relaxation" there is an avoided crossing between the energetically close to each other surface states with the same spin projection: the branches of the Dirac cone (stemming from the surface QL block) and the Rashba-type state (from the subsurface SL block). The branches with a positive/negative spin projection (red/blue and pink/cyan colors) repel each other and due to such hybridization, the continuity of the upper branches of the cone becomes violated by the presence of an energy gap. The size of this gap is determined by the degree of the wave function overlap of the Dirac and Rashba states. In the present Bi$_2$Te$_2$Se/PbBi$_2$Te$_2$Se$_2$ superlattice, the intersection occurs at small $k_\|$ and the Rashba-type state smoothly transforms to the Dirac one.

The additional surface relaxation of the studied slabs does not lead to any noticeable changes neither in the structure nor band dispersion. For this reason, we treat it as an insignificant contribution and do not include into consideration for all QL/SL superlattices presented below.

The reason for the appearance of the Rashba state resided in the second vdW block can be understood from consideration of the surface electrostatic potential. The calculations show a positive bending of the electrostatic potential ($\Delta V > 0$), which decays up to the third block from the surface of the slab (Fig.~\ref{fig1}(f)). Here $\Delta V = V_\mathrm{slab} - V_\mathrm{bulk}$ along $z$ axis normal to the surface, where $V_\mathrm{slab}$ and $ V_\mathrm{bulk}$ are the slab and bulk potentials, respectively, averaged over the $xy$ plane. To obtain the bulk potential $V_{\mathrm{bulk}}$ we choose the central part of the slab that has the bulk-like potential, then extend it on the whole slab. Since there are no dangling bonds on the surface aligned with the vdW plane, the inducing of a positive potential bending can be explained by the presence of the TSS inherent in 3D TI at the boundary with vacuum. This leads to the localization of a negative charge near the surface. As can be seen from the figure, the maximum of $\Delta V \approx 0.2$~eV is located exactly at the surface layer. On the spectra the non-zero potential bending in the second block as well as an inequivalence of QL and SL blocks are reflected in the emerging of the above-described spin-polarized states of the Rashba type.

One can highlight an important role of interlayer distances within the vdW blocks in formation of the potential bending. Namely, changes in the first atomic layers to the surface can drastically  modify the potential profile. Together with a change in the position of bulk bands, it can lead to hybridization of the surface states and rearrangement of TSS on the 3D TI surface. Similar effects of hybridization of surface states from different structural blocks also arise in the previously studied topologically nontrivial superlattice structures Sn(Bi,Sb)$_{4}$Te$_7$~\cite{vergniory2013bulk,vergniory2015electronic}, PbBi$_{6}$Te$_{10}$~\cite{papagno2016multiple}.

Alternatively, the TB model of the surface constructed on the bulk wave functions does not take into an account the potential bending caused by the surface. As shown on Fig.~\ref{fig1}(g), within this model the spectrum of the QL termination demonstrates the hybridization-induced gap in the Dirac cone branches when they intersect the topmost bulk band, whereas the Rashba state does not appear explicitly being merged with the bulk states. In order to be well-defined and split off from the bulk valence states, it needs an enough slowly-decreasing surface potential. As shown above, the Dirac state residing on the surface is responsible for inducing of such potential.

The revealed effect of relatively small changes in interlayer distances (of a relaxation origin in the above-considered case) on the electronic structure of surface states, especially noticeable for QL termination, gives us a hint that the same kind of changes can be caused by variations in the atomic sizes in composition. Below, by substitution of the particular atomic layers with isovalent atoms we consider other hypothetical QL/SL superlattices with the same crystal structure to analyze the generalization for arising of the aforementioned surface states and influence of the electrostatic potential.

\begin{figure*}[t]
\begin{center}
\includegraphics[width=\textwidth]{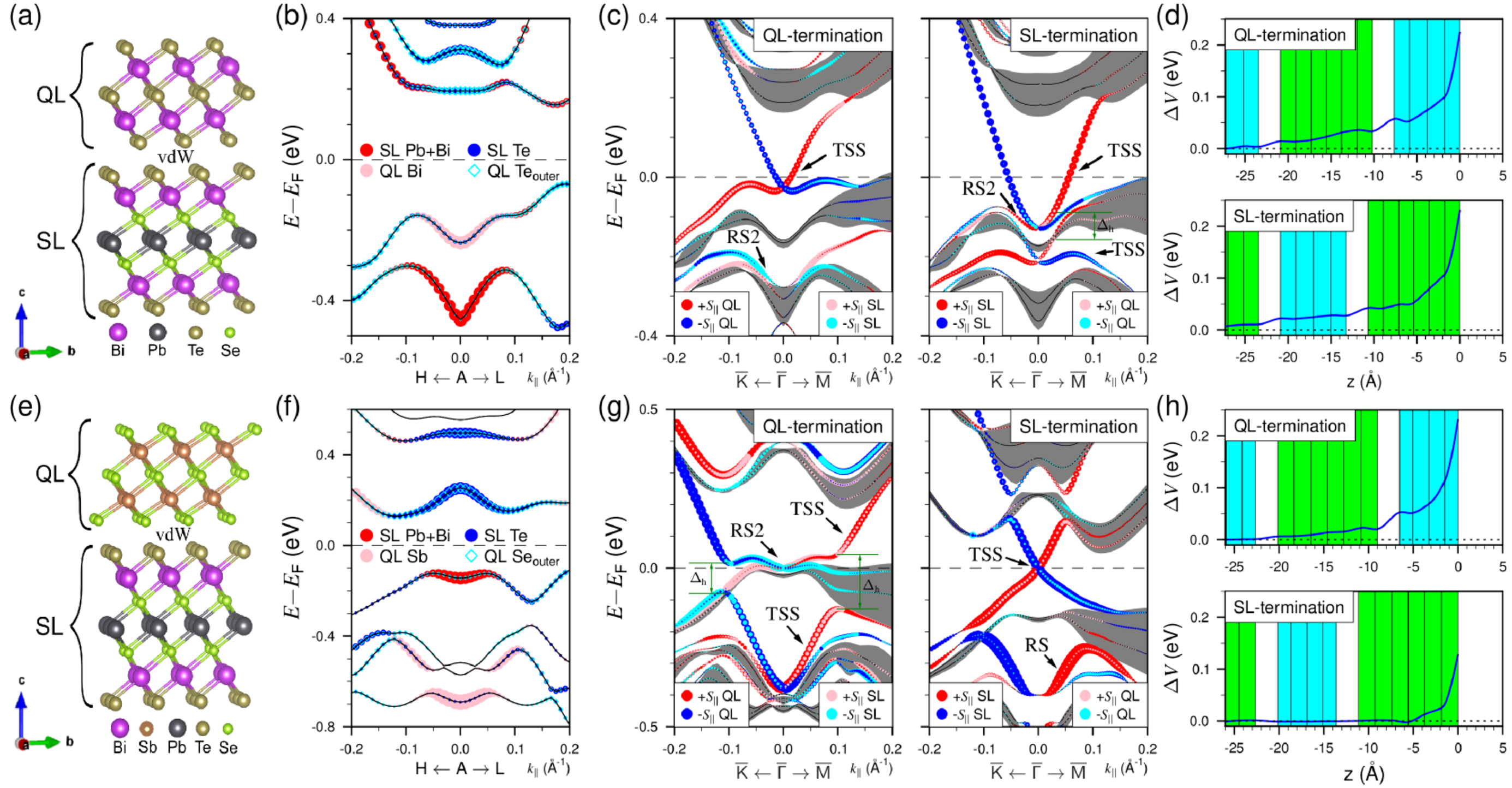}

\caption{(a, e) Crystal structure; (b, f) orbital-resolved bulk and (c, g) in-plane spin-resolved surface band structures; (d, h) potential bending profile near the surface of the topologically nontrivial (a-d) Bi$_{2}$Te$_{3}$/PbBi$_{2}$Te$_{2}$Se$_{2}$ and (e-h) Sb$_{2}$Se$_{3}$/PbBi$_{2}$Te$_{2}$Se$_{2}$ superlattices.}\label{fig2}
\end{center}
\end{figure*}

On the base of the considered Bi$_{2}$Te$_{2}$Se/PbBi$_{2}$Te$_{2}$Se$_{2}$ vdW superlattice, by replacing Se atoms with heavier Te atoms in the central layer of QL, one can obtain the Bi$_{2}$Te$_{3}$/PbBi$_{2}$Te$_{2}$Se$_{2}$ compound (Fig.~\ref{fig2}(a)). The calculated topological invariants $(\nu_0; \nu_1 \nu_2 \nu_3)=(1;001)$, where $\nu_0$ is the strong
$\mathbb{Z}_2$ index, and $\nu_1$, $\nu_2$, and $\nu_3$ are the weak
$\mathbb{Z}_2$ indices, show that the electronic structure possesses topologically nontrivial phase. The band gap in Bi$_2$Te$_3$/PbBi$_2$Te$_2$Se$_2$ is much smaller ($\approx 150$~meV), see Fig.~\ref{fig2}(b), than in the parent superlattice ($\approx 270$~meV). The replacement by heavier atoms in the QL block also results in a shift of the valence band contributed by orbitals of the QL-block atoms upward in energy. The reason for that is an increase of the inner  Bi--(Se$~$$\rightarrow$$~$Te) interlayer distance in the QL block by $\sim 12 \%$ (from 1.83~\AA~to 2.05~\AA) and probably a higher core potential. In this case, the first and second bulk valence bands are reversed near the A point with respect to their order in the Bi$_2$Te$_2$Se/PbBi$_2$Te$_2$Se$_2$. Therefore, the hybridization of the Dirac cone with the resided in the second vdW block Rashba state within the principal band gap at $E \sim -0.15$~eV occurs on the SL termination (panel (c) for SL termination), leading to the gap $\Delta_{\rm h}\approx$ 70 meV. The potential bending profile for both terminations is similar (panel (d)) and it vanishes at the third block as well. On the QL termination the RS2 state is mixed with bulk states in close vicinity of \Gammapoint. However, with an increase of a number of blocks it turns into a well-defined Rashba state with doubly degenerate branches at~\Gammapoint.

Keeping the SL block the same, we choose Sb$_2$Se$_3$ as the QL block for constructing Sb$_2$Se$_3$/PbBi$_2$Te$_2$Se$_2$ superlattice (Fig.~\ref{fig2}(e)). Note that Sb$_2$Se$_3$ does not exist in the QL-structured trigonal bulk phase, however, under applied tensile strain ($\sim 7$~\%) it can be achieved~\cite{Sb2Se3_strain}. The low-energy valence bands in the {Sb$_2$Se$_3$/PbBi$_2$Te$_2$Se$_2$ bulk spectrum are a little more intricate (Fig.~\ref{fig2} (f)) due to more prominent orbitals of the inner Se atoms. SOC remains large enough to guarantee the same nontrivial $\mathbb Z_2$  class as in Bi$_2$Te$_3$/PbBi$_2$Te$_2$Se$_2$. Two features can be noted for this particular superlattice. First, the spectrum of the SL termination (panel (g)) exhibits surface states resided only in the surface (SL) block (all states are marked in blue and red according to negative/positive sign of their spin polarization). This is a result of the surface potential behavior -- its bending vanishes in the center of the first block (panel (h) for SL termination), for this reason we do not observe the surface states primarily localised in the second block. Second, in the case of the QL termination one can clearly see two shifted parabola-like dispersed surface states localized in the second (SL) block (electronic states in cyan and pink at $E=0$ eV), which is the characteristic feature of the Rashba state. Since the intersection with the topological state occurs at large $k_\|$ ($\approx 0.09$~\AA $^{-1}$)}, the branches manage to change the sign of a slope. On the other hand, the slope of the topological state remains almost unchanged when approaching the bulk valence band. The hybridization gap, $\Delta_{\rm h}$, varies from 0.1 eV along the \GammaK~direction to 0.17 eV along the \GammaM~direction due to different width of the bulk band.

\begin{figure*}
\begin{center}
\includegraphics[width=\textwidth]{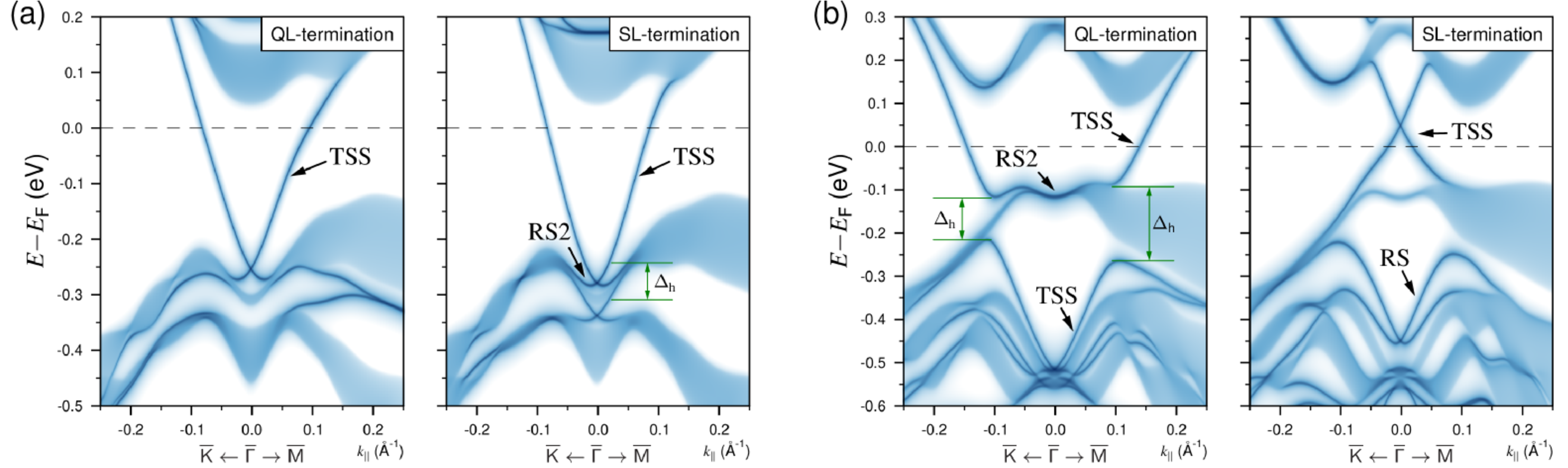}

\caption{Tight-binding surface spectra of (a) Bi$_{2}$Te$_{3}$/PbBi$_{2}$Te$_{2}$Se$_{2}$ and (b) Sb$_{2}$Se$_{3}$/PbBi$_{2}$Te$_{2}$Se$_{2}$ superlattices.}\label{fig2b}
\end{center}
\end{figure*}

\begin{figure*}
\begin{center}
\includegraphics[width=0.8\textwidth]{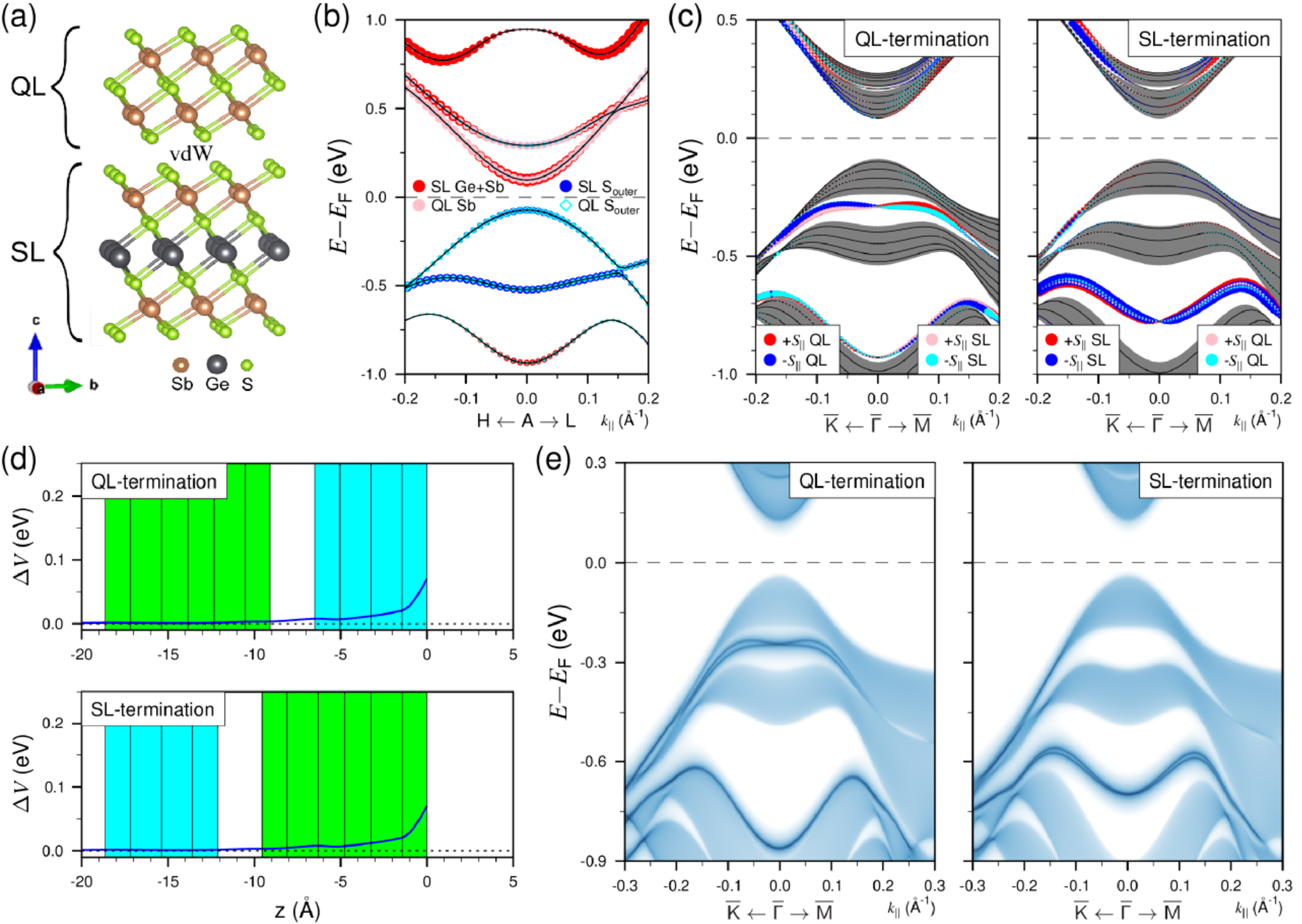}
\caption{(a) Crystal structure, DFT (b) orbital-resolved bulk and (c) spin-resolved surface electronic structures of the topologically trivial Sb$_2$S$_3$/GeSb$_2$S$_4$ superlattice. (d) Potential bending profile near the surface. (e) Tight-binding surface spectra.}
 \label{fig3}
\end{center}
\end{figure*}

The reason for the difference in the potential bending between the QL and SL terminations of Sb$_2$Se$_3$/PbBi$_2$Te$_2$Se$_2$, in contrast to  Bi$_2$Te$_3$/PbBi$_2$Te$_2$Se$_2$, can be related to that the outermost layers of the QL and SL blocks are now different (Te--Bi-- ... and Se--Sb--..., respectively, instead of the identical Te--Bi--...). However, the behavior of $\Delta V$ inside the slab cannot be predicted, since there is a combined effect of the chemical composition of atomic layers and the distances between them, which determines the charge redistribution. Up to this point, all the considered cases are related to the presence of a TSS, negatively charging the surface and causing the potential bending near the surface in such vdW structures, which are free of dangling bonds. TB calculations (Fig.~\ref{fig2b}) prove the same idea -- the Rashba states appear even without electrostatic potential as states merging with the bulk continuum (see SL termination in (a) and QL termination in (b)). The potential, in turn, squeezes these surface states out from the bulk band.

Consistent with the examples analyzed above, the question naturally arises of how the potential bending will behave and, hence, how the surface states will manifest in the QL/SL vdW structures in the absence of a TSS on the surface. To this end we constructed the hypothetical structures with isovalent but more lighter atoms like Ge, Sb, S. With this choice of atoms, as an example, the structure Sb$_2$S$_3$/GeSb$_2$S$_4$ was studied (Fig.~\ref{fig3}(a)). With such chemical composition the inversion of the bands that form the band gap no longer occurs (Fig.~\ref{fig3}(b)) and  related $\mathbb Z_2$ topological invariants are equal to zero. The spectrum of both surface terminations does not contain any states within narrow bulk band gap  and only trivial surface states arise in the local band gaps (Fig.~\ref{fig3}(c)). In the absence of the TSS a small rapidly decreasing positive bending of the potential emerges near the surface (Fig.~\ref{fig3}(d)), the magnitude of which in the surface layer is approximately 5 times less than in the topologically nontrivial superlattices. For this reason, the TB-based surface spectra (Fig.~\ref{fig3}(e)) completely reproduce the DFT slab spectra.

Summarizing these observations, for superlattice structures consisting of inequivalent structural blocks and being 3D TIs, in addition to typical for this class of materials the Dirac surface state, there is a tendency for the appearance of another spin-polarized surface state of the Rashba type near the Fermi level. This state is split off from the bulk valence band and formed by atomic orbitals from the second (subsurface) structural block. That is, for different terminations the Rashba state is split off from the first or the second bulk valence band. Moreover, a presence of the symmetry-protected Dirac state contributes to the Rashba state development through the induced electrostatic potential. With that the bulk bands relative shift can lead to a hybridization of the surface states from different blocks and significant rearrangement of the surface electronic structure near the Fermi level. The latter is very sensitive to changes in the composition and interlayer distances within the QL(SL) structural blocks (which in turn depend on the sizes of the constituent atoms). It should be mentioned that the described mechanism is intrinsic and is not related to the structural defects.


\section{$\mathrm{Mn(Bi_{1-x}Sb_{x})_{4}Te_{7}}$ QL/SL superlattice}

 \label{section_4}

When considering non-magnetic QL/SL superlattices, it was discovered that the surface potential affected by the topological Dirac state can lead to the appearance of an additional Rashba state in the energy gap region, which demonstrates complex hybridization with the Dirac state itself. In the group of extensively studied magnetic topological systems there are series (MnBi(Sb)$_2$Te$_4$)$\cdot$(Bi(Sb)$_2$Te$_3$)$_n$ vdW compounds composed of alternating magnetic SL and non-magnetic QL blocks among which MnBi(Sb)$_4$Te$_7$ are the simplest examples ($n=1$) which are isostructural to the discussed above non-magnetic QL/SL superlattices. In this regard one can expect that the effect of surface potential should manifest itself in magnetic superlattices too. Indeed, the electronic structure of pure MnBi$_4$Te$_7$ and MnSb$_4$Te$_7$ AFM topological insulators is well studied \cite{rienks2019large,klimovskikh2020tunable,eremeev2021topological,He2020,Xu_NatComm2022,Liu_PRX2021,Huan_PRL2021}.
However, in contrast to naturally $n(p)$-doped Bi(Sb)-based compounds, the charge neutral Mn(Bi$_{1-x}$Sb$_{x}$)$_4$Te$_7$ system is more attractive from the point of view of surface conductivity. It is noteworthy that in  Bi$_{1-x}$Sb$_{x}$-based systems a dependence of magnetic state on concentration and distribution of anti-site defects was revealed~\cite{Xi_2024}. Therefore, large concentration of Mn atoms in Bi$_{1-x}$Sb$_{x}$ sublattice can turn AFM magnetic state inherent in the parent compounds to FM one. Neglecting the anti-site defects here we consider AFM and FM phases of Mn(Bi$_{1-x}$Sb$_{x}$)$_4$Te$_7$ within virtual crystal approximation with the main focus on $x=0.3$ case that corresponds to the charge-neutrality point in the experiment \cite{Chen_PRB2021}. We also compare these results with those for $x=0.5$ (50/50 Bi-Sb intermixing). Within this approach, only changes in lattice parameters and interlayer distances are taken into account, which, as shown above, can significantly affect the dispersion of the emerging surface states. Since Mn anti-site defects are not taken into account, Mn(Bi$_{1-x}$Sb$_{x}$)$_4$Te$_7$ keeps the interlayer AFM ground state with the energy difference of $\approx 0.2$~meV per Mn pair between AFM and FM configurations that is of the same order as in parent MnBi$_4$Te$_7$ and MnSb$_4$Te$_7$. In the FM phase, following the parent compounds, it is characterized by $\mathbb{Z}_4=2$ topological invariant possessing the FM axion insulator phase.

For calculation of the surface electronic structure of SL termination of Mn(Bi$_{1-x}$Sb$_{x}$)$_4$Te$_7$ within AFM state we use the same slab thickness as for non-magnetic superlattices, however, for QL termination the slab is thickened so that it contains the same four magnetic blocks to provide a compensated antiferromagnet state (zero total magnetic moment), see Fig.~\ref{fig_MBST}(a). The same geometry of the slab with QL termination is utilized for FM calculations. For SL termination in the FM phase, the final choice is set to 55 layers, which is by 12 layers (QL+SL) more than in the AFM case (Fig.~\ref{fig_MBST}(a)). As will be seen below, even such thicknesses are not completely enough for description of the surface band structure in the FM phase, but we are limited by computational facilities.

\begin{figure*}
\begin{center}
\includegraphics[width=\textwidth]{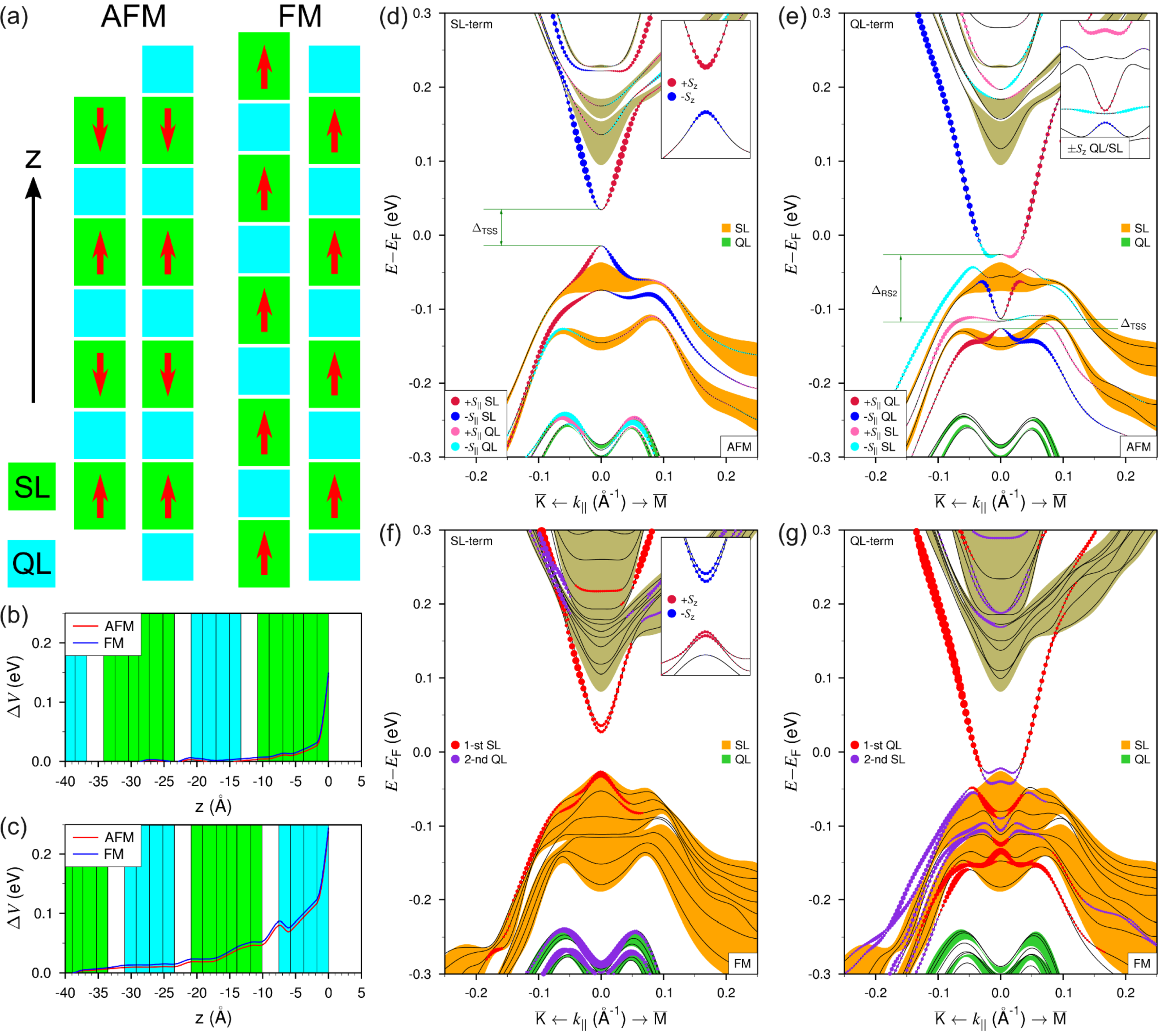}
\caption{(a) Schematic structure of Mn(Bi$_{0.7}$Sb$_{0.3}$)$_4$Te$_7$ slabs used for electronic structure calculations of SL- and QL-terminated surfaces in AFM and FM magnetic states with red arrows indicating spin direction on Mn atoms. Surface potential bending $\Delta V(z)$ at (b) SL- and (c) QL-terminated surfaces. Red and blue lines correspond to AFM and FM order in the slab.  Spin-resolved surface band spectra of (d) SL- and (e) QL-terminated surfaces in AFM state. Dark/light dots indicate in-plane ($S_\|$) positive (red) and negative (blue) spin components of the states localized in outer/second vdW block. Block-resolved surface band spectra of (f) SL- and (g) QL-terminated surfaces of the FM Mn(Bi$_{0.7}$Sb$_{0.3}$)$_4$Te$_7$ with red and violet dots for the states predominantly localized in the outermost and second vdW blocks.
Shaded areas in (d--g) mark bulk states projected onto (0001) surface. Insets in (d-f) show out-of-plane ($S_z$) spin component near the gap in the Dirac state.
}
 \label{fig_MBST}
\end{center}
\end{figure*}

As can be seen in Figs.~\ref{fig_MBST}(b,c) the behavior of the surface potential shows similarities with that found for non-magnetic QL/SL topological insulator superlattices: $\Delta V$ is rapidly decaying within the surface block on the SL termination while it is propagating much deeper, demonstrating sizable potential bending in the subsurface block on the QL termination. It is noteworthy that despite the fact that the atomic structure of the vdW blocks in our calculation is identical in the AFM and FM states, the difference in $\Delta V$ is noticeable and the latter is systematically higher.

Similar to non-magnetic QL/SL counterparts and parent magnetic Bi- and Sb-based superlattices the top two bulk valence bands are predominantly contributed by orbitals of atoms of different blocks: SL (the topmost) and QL (the second one). However, each of bulk band in the magnetic superlattices is magnetically split into two subbands. Projections of these subbands onto (0001) surface are highlighted by different colors in Figs.~\ref{fig_MBST}(d-g). Like in parent MnBi$_4$Te$_7$ and MnSb$_4$Te$_7$ AFMTIs the splitting between subbands of the magnetic SL block is much larger than that for QL localized subbands (Figs.~\ref{fig_MBST}(d-e)).

The spin-helical Dirac band of the SL termination of Mn(Bi$_{0.7}$Sb$_{0.3}$)$_4$Te$_7$ in AFM state is located within the inverted band gap of $\approx 130$~meV width and shows the gap $\Delta_{\rm TSS} = 48.6$~meV at the $\bar\Gamma$ point (Fig.~\ref{fig_MBST}(d)).
This exchange gap in the Dirac state is in between those in parent Bi- (70 meV \cite{klimovskikh2020tunable}) and Sb-based (32.5 meV \cite{eremeev2021topological}) QL/SL superlattices and it is smaller, 41.1 meV, when the Sb concentration $x$ increases to 0.5 (50/50 Sb-Bi intermixing case).
In the vicinity of the magnetic exchange induced gap the out-of-plane spin component arises in the Dirac state, see Fig.~\ref{fig_MBST}(d, inset) where $S_z$ spin component of the Dirac state resided on the top surface of the slab are shown. It is of the opposite sign for lower and upper branches. Accordingly, the Dirac states localized on the bottom surface of the slab have opposite directions of $S_z$ as well as $S_\|$ spin components. With the chosen setup for AFM slab when spins on the Mn atoms of the bottom and top SLs are pointed inward, the lower(upper) branches of the gapped Dirac state demonstrate negative(positive) sign of $S_z$. The sign would be inverted for alternative setup of the AFM order in the slab, when the Mn magnetic moments will be opposite and hence pointed outward in the surface SL blocks. The Rashba state on the SL termination resides near the edge of the QL-derived bulk subbands and demonstrates only a tiny exchange gap.

On the QL termination of the AFM Mn(Bi$_{0.7}$Sb$_{0.3}$)$_4$Te$_7$, like in non-magnetic counterparts, see e.g. Bi$_2$Te$_2$Se/PbBi$_2$Te$_2$Se$_2$ (Figs.~\ref{fig1}(d,e)), the Rashba state splits off from the SL-derived bulk band, however, being localized in the magnetic block it acquires a huge exchange gap $\Delta_{\rm RS2} = 90.6$~meV at $\bar\Gamma$ (Fig.~\ref{fig_MBST}(e)) with sizeable $S_z$ spin component in its vicinity (inset). Note that outer and inner branches of the Rashba state are split off from the first and second bulk SL-subbands.
The same as on QL terminations in non-magnetic superlattices the Dirac state penetrates deep in the bulk continuum and hybridizes with the emerging Rashba state. Since on the QL termination the Dirac state predominantly resides in the surface non-magnetic block, only partially penetrating the subsurface magnetic SL, the exchange splitting $\Delta_{\rm TSS}$ is much smaller, only 12.8~meV as well as the values of $S_z$ at the Dirac gap edges are substantially reduced (Fig.~\ref{fig_MBST}(e), inset). In this case, at the $\bar \Gamma$ point the gapped Dirac state appears in the gap between the bulk SL-subbands and the inner branch of the Rashba state with lifted degeneracy lies within the Dirac state gap.
Note that, in contrast to SL-terminated surface, where the $\Delta_{\rm TSS}$ decreases with increase in $x$, on the QL termination of Mn(Bi$_{0.5}$Sb$_{0.5}$)$_4$Te$_7$ it slightly increases to 15.6 meV.

Lets turn to the FM phase of the Mn(Bi$_{1-x}$Sb$_{x}$)$_4$Te$_7$ QL/SL superlattices. The surface spectrum of ferromagnetic SL-terminated Mn(Bi$_{0.7}$Sb$_{0.3}$)$_4$Te$_7$ is shown in Fig.~\ref{fig_MBST}(f), where states localized in the first two blocks on the top and bottom surfaces are highlighted with different colors. In this case, first, the topmost SL-localized bulk valence band subbands become broader so the gap between them almost disappears. Second, the Dirac states localized on the top and bottom surfaces are no longer degenerate. This is due to
the fact the Mn spin magnetic moments in the bottom and top SLs have opposite directions, inward and outward, respectively (Fig.~\ref{fig_MBST}(a)). Hence, the gapped Dirac states localized on opposite surfaces acquire the same sign of the $S_z$ spin components (Fig~\ref{fig_MBST}(f), inset) and repel each other. Thus, the splitting in the gapped Dirac cone branches at $\bar\Gamma$, of 4.2/8.3 meV in lower/upper branches, respectively, is due to the finite slab thickness. This splitting is much larger (11.2/17.5 meV) in the slab of 43-layer thickness, the same as in the AFM SL-terminated case, and hence one can expect its rapid diminishing with further increase in the slab thickness. Besides the change in the spin texture in the gapped Dirac state, the change in the magnetic state from AFM to FM also results in increase of the $\bar\Gamma$ gap that is about 63 meV in the ferromagnetic case.
As regards the Rashba state on the SL termination in the FM phase, it resides at the edge of deep-lying QL-derived bulk bands like in the AFM case.

In the QL-terminated surface spectrum of FM superlattice (Fig.~\ref{fig_MBST}(g)) the finite slab thickness affects the emerging Rashba state residing in the subsurface magnetic SLs. It is due to the same reason as for the Dirac state on the SL termination -- the states localized on opposite sides of the slab carry $S_z$ spin of the same sign. Undoubtedly, this artificial splitting will also decrease with increasing the slab thickness. The $\bar\Gamma$ gap of $\approx 9.5$~meV in the Dirac state, resided mainly in the surface non-magnetic QL block, is almost the same as in the AFM case at this surface termination and does not show sizable artifact finite thickness splitting. However, in contrast to AFM phase, where it resides in the gap between the topmost valence band subbands, in the FM case the Dirac state falls into the region of the bulk continuum.

\section{Conclusion} \label{section_5}

Surface electronic structure of the bulk superlattices, consisting of the pnictogen chalcogenide vdW QL and SL structural elements, which individually form 3D TI crystals, have been studied. Surface of such vdW structures has two types of termination: QL or SL block. Based on a systematic study of the orbital composition and spin structure, it was established that depending on the termination in addition to the Dirac cone state localized in the surface block, the spin-split state specific to such superlattice structures emerges near the Fermi level. Namely, it is a surface spin-polarized state of the Rashba type, which splits off from the bulk valence band formed by the orbitals of atoms from the second structural block. The reason for the appearance of this state is an inequivalence of the surface and subsurface structural blocks and a slowly-decaying bending of the electrostatic potential near the surface. In energy spectra, when such a trivial state becomes energetically close to  a topological one, their hybridization occurs, accompanied by repulsion of spectral branches with the same spin projection. This leads to a splitting of the Dirac cone and formation of a more complex electronic structure near the Fermi level relative to the surface spectra of TIs formed solely by QL or SL blocks. Considering a number of isostructural hypothetical compounds, including topologically trivial ones, the important role of the presence of the topological state in inducing the surface potential bending was disclosed. Moreover, it was found that the composition and parameters of the crystal structure significantly affect the localization and magnitude of the bending, as well as the relative position of the bulk bands. Together, they determine the features of the electronic structure of the surface.
All effects found for non-magnetic QL/SL superlattices, related to different behavior of the surface potential at different terminations, are also preserved for their magnetic counterparts Mn(Bi$_{1-x}$Sb$_{x}$)$_4$Te$_7$ both in antiferro- and ferromagnetic states that support AFM TI and FM axion insulator topological phases in the superlattice and in which exchange splitting occurs in the Dirac as well as in the emergent Rashba states.
The revealed mechanism can be applied to interpret experimental observations in numerous real non-magnetic and magnetic structures of the QL/SL type.

\acknowledgments{I.A.Sh. gratefully acknowledges financial support from the Ministry of Education and Science of the Russian Federation within State Task No. FSWM-2020-0033 (in the part of calculation of non-magnetic superlattices). S.V.E. acknowledges support from the Government research assignment for ISPMS SB RAS, project FWRW-2022-0001 (in the part of calculation of magnetic superlattices). E.V.C. acknowledges support from Saint Petersburg State University (Project ID No. 95442847).}

\end{document}